\documentclass[aps,amsmath,amssymb,amsfonts,floatfix,reprint]{revtex4-1}
\usepackage{inputenc}%[utf8x]
\usepackage[pdftex]{graphicx}
\usepackage[squaren]{SIunits} %\setstretch{1.5}
\usepackage{color}
\usepackage{appendix}

\begin{document}

\title{Dependence of transverse magneto-thermoelectric effects on inhomogeneous magnetic fields}

\author{A.S. Shestakov$^{1}$}
\author{M. Schmid$^{1}$}
\author{D. Meier$^{2}$}
\author{T. Kuschel$^{2}$}
\author{C.H. Back$^{1}$}

\affiliation{$^{1}$Institute of Experimental and Applied Physics, University of Regensburg, 93040 Regensburg, Germany}
\affiliation{$^{2}$Center for Spinelectronic Materials and Devices, Physics Department, Bielefeld University, 33615 Bielefeld, Germany}
\date{\today}

\begin{abstract}
Transverse magneto-thermoelectric effects are studied in permalloy thin films grown on MgO substrates. We find that small parasitic magnetic fields below 1 Oe can produce artifacts of the order of 1$\%$ of the amplitude of the anisotropic magneto-thermopower which is also detected in the experiments. The measured artifacts reveal a new source of uncertainties for the detection of the transverse spin Seebeck effect. Taking these results into account we conclude that the contribution of the transverse spin Seebeck effect to the detected voltages is below the noise level of 20 nV.

\end{abstract}

\maketitle

\section{Introduction}
The rapidly evolving field of spincaloritronics \cite{timo1,timo2} has attracted a lot of attention in the past years. Particularly, the transverse spin Seebeck effect (TSSE), at first sight a seemingly simple effect to measure, has been the target of many experimentalists. 
However, it turned out that the TSSE \cite{tsse1,tsse2,tsse3,tsse4,tsse5,tsse6,tsse7,tsse8,tsse9,timo3,timo4,timo5,prlmax,AMTEP}, especially when using metallic ferromagnets \cite{tsse1,tsse2,tsse3,tsse4,tsse5,timo4,prlmax,AMTEP}, is under strong discussion. Other magneto-thermoelectric effects, like the anomalous Nernst effect (ANE) \cite{timoANE,tsse4,tsse5,prlmax,AMTEP} and the anisotropic magneto-thermopower (AMTEP, also known as the planar Nernst effect (PNE)) \cite{tsse2,prlmax,AMTEP,AMTEP2,AMTEP3} play an important role in the TSSE measurements and contribute to the detected signals. ANE and TSSE signals are antisymmetric with respect to the external magnetic field, have the shape of hysteresis loops and both follow the $\cos \Theta_{0}$ dependence, where $\Theta_{0}$ is the angle between the magnetization vector $\vec{M}$ and the temperature gradient $\vec{\nabla} T$. The AMTEP can be identified by its $\sin 2\Theta_{0}$ angular dependence and symmetric shape.
Thus, it was postulated that the ANE caused by an unintended out-of-plane temperature gradient (due to heat flux into surrounding area by radiation \cite{prlmax} and through electrical contacts by thermal conductivity \cite{AMTEP}) is the main effect masking TSSE in metallic ferromagnets. Small static environmental magnetic fields of the Earth magnetic field's order of magnitude and lower are typically not considered in experimental set-ups. However, in nanovolt range signal measurements this fields could lead to unexpected effects those are mask or mimic the investigated effect. In this article we show that the AMTEP generated by solely in-plane $\vec{\nabla} T$ in presence of very small static parasitic magnetic field can produce an significant antisymmetric artifact that shares the $\cos \Theta_{0}$ dependence with the ANE and TSSE. Therefore, the third source of uncertainties is found for TSSE experiments which then could lead to wrong interpretations of the observed data. 
%This fact gives us an explanation on the question that was left open in our previous work \cite{prlmax}.

\begin{figure}
	\centering
		\includegraphics[width=8.5cm]{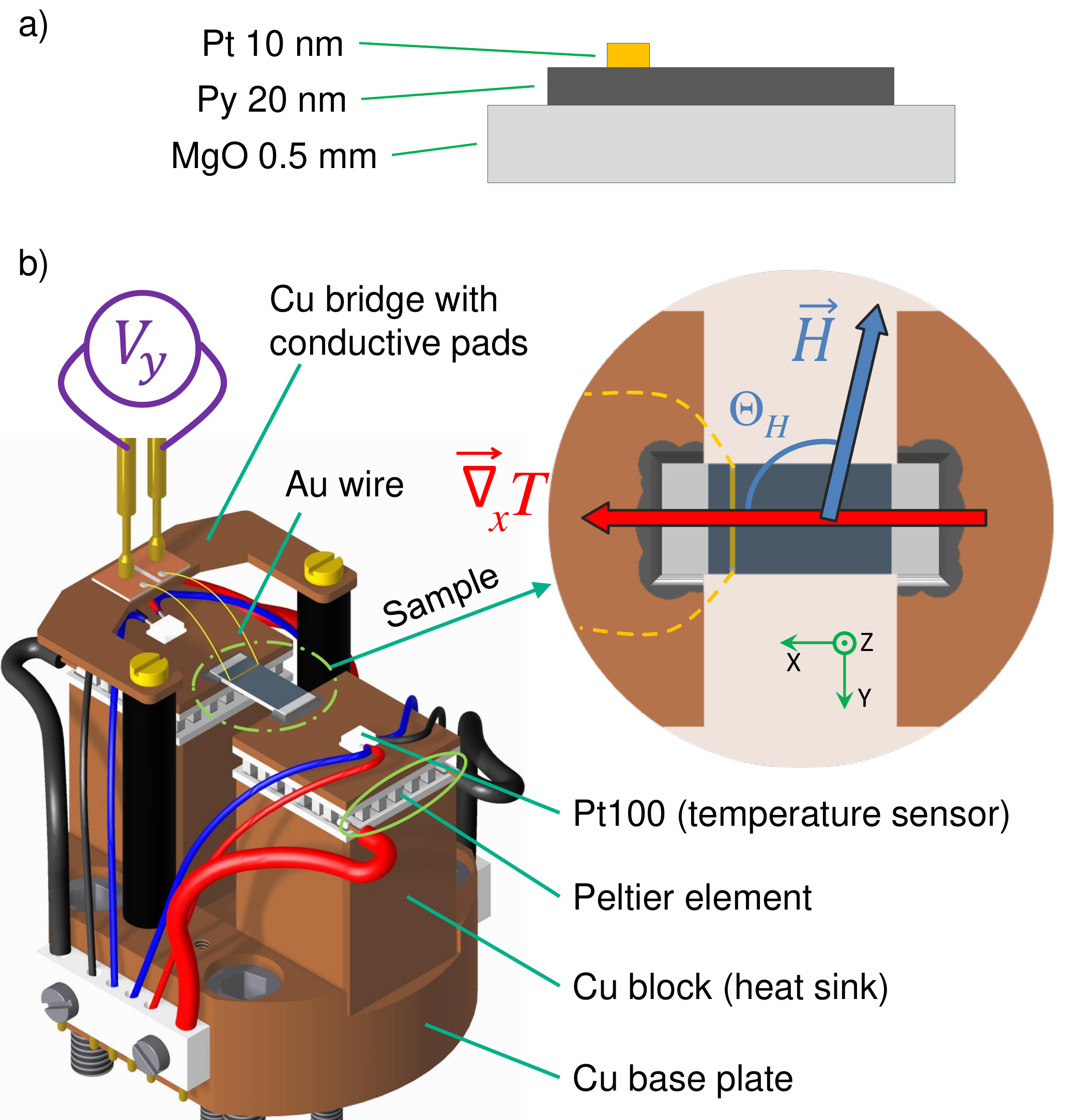}
		\caption{a) Sample parameters: 0.5 mm thick MgO substrate (area of 10 mm $\times$ 4 mm), 20 nm thick Py film (area of 6 mm $\times$ 4 mm) on top of the substrate and 10 nm thick Pt strip (0.1 mm $\times$ 4 mm) grown on the Py film. 
		b) Sample holder: Two Cu blocks are attached to a common Cu base plate, and carry the Peltier elements with Pt100 temperature sensors attached to them (these sensors measure the temperatures $T_{1}$ and $T_{2}$, see Appendix \ref{app:temp}). Close-up: The sample is connected by thermally conductive glue to the Cu interfaces of the Peltier elements, which produce the temperature gradient $\vec{\nabla}_{x} T$. The external magnetic field $H=-44..+44$ Oe is applied in the x-y plane with the angle $\Theta_{H}$ in reference to the x-axis.}
		\label{sample}
\end{figure}

\section{Sample properties and experimental set-up}
A 20 nm thick permalloy film (fig.\,\ref{sample}a) with area of 6 mm $\times$ 4 mm is deposited onto a $0.5$ mm thick MgO substrate (area of 10 mm $\times$ 4 mm) using sputter deposition. To detect the expected pure spin current via the inverse spin Hall effect \cite{ishe1,timo6,ishe2} a 10 nm thick Pt strip is deposited in-situ through a shadow mask onto the Py film.

In fig.\,\ref{sample}b the sample holder is shown. A Cu base plate holds two Cu blocks (heat sinks). Peltier elements are attached to each Cu block. The sample is glued onto the Cu pads, which are mounted as heat sinks onto the Peltier elements. The Peltier elements provide a controlled temperature gradient in the sample plane along the x-axis (see close-up in the inset in fig.\,\ref{sample}b). The temperatures of the Peltier elements are measured with Pt100 temperature sensors. The temperature value at the position of the Pt strip  $T_{str}$ is used as the base temperature. For details concerning the determination of $T_{str}$ and the temperature difference on the sample $\Delta T_{x}$ see Appendix \ref{app:temp}. 
 20 $\mu$m thin Au wires are glued to the ends of the Pt strip detector to provide electrical contacts to a nanovoltmeter. Opposite ends of the Au wires are glued to conductive pads of a Cu bridge, where a voltage signal gets picked up by two spring contacts. The conductive pads are electrically isolated one from another and from the Cu bridge. 

\begin{figure}
	\centering
		\includegraphics[width=8.5cm]{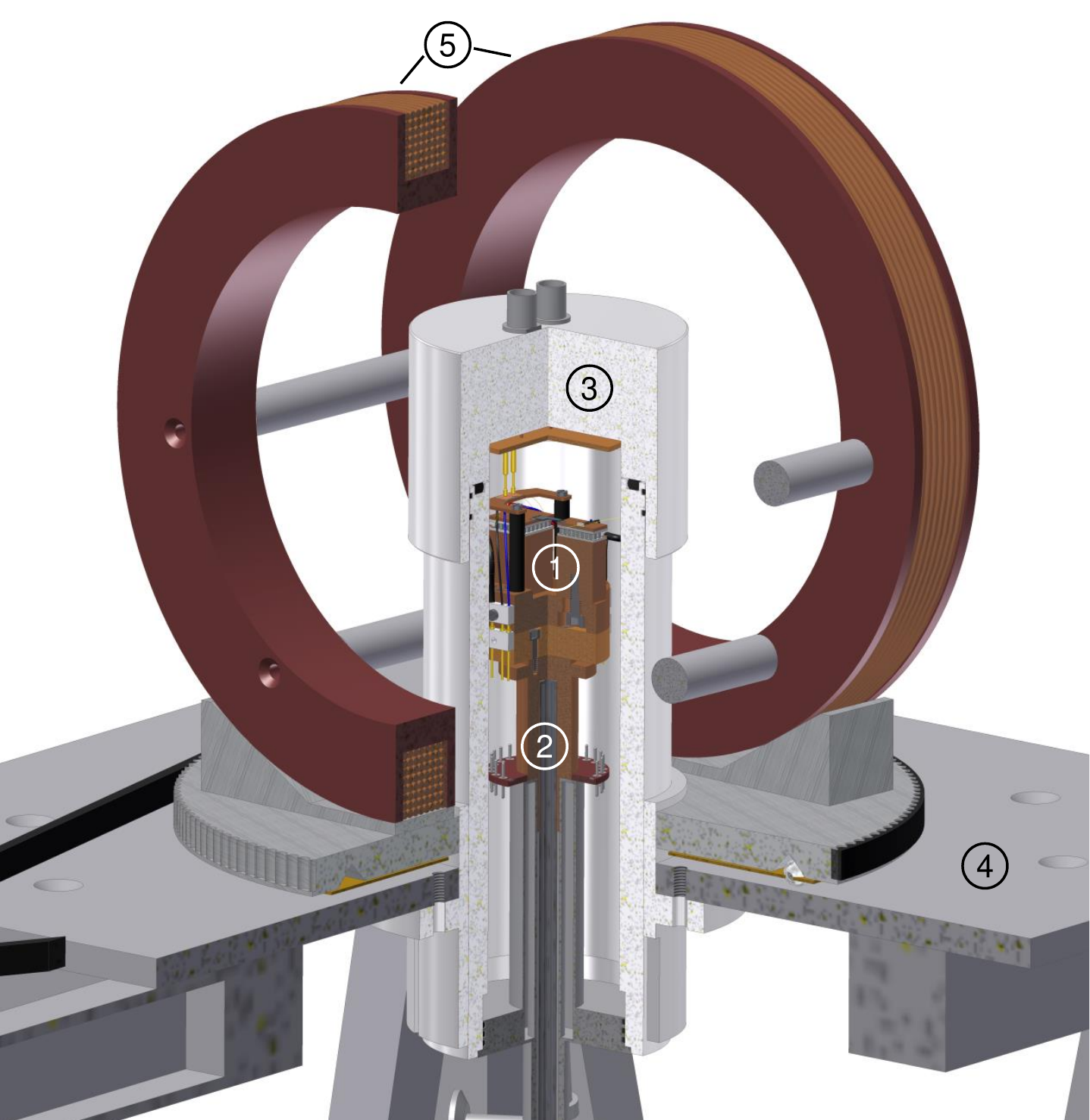}
		\caption{Set-up: The sample holder (1) rests on a pillar (2) and is placed in the middle of the vacuum chamber (3) without contact to its walls. The vacuum chamber is mounted onto a table (4) and fixed rigidly. Helmholtz coils (5) surround the vacuum chamber and are able to rotate around it. Additionally, an Al shield (not shown here) covers the whole set-up. This shield is used to decrease external temperature variations and electro-magnetic noise during the measurements.}
		\label{setup}
\end{figure}

Measurements are performed in a vacuum chamber (see fig.\,\ref{setup}) at a base pressure $2 \cdot 10^{-6}$ mbar to prevent influence of convection and thermal conductivity of air. The external magnetic field with values between $H=-44..+44$ Oe is produced by Helmholtz coils, which can be rotated around the fixed measuring chamber. The magnetic field is applied in the x-y plane of the sample with angles in steps of  $30^\circ$  ($\Theta_{H}=0^\circ,30^\circ,..,360^\circ$, see close-up in fig.\,\ref{sample}b) with respect to the x-axis.

\begin{figure*}
	\centering
		\includegraphics[width=17.7cm]{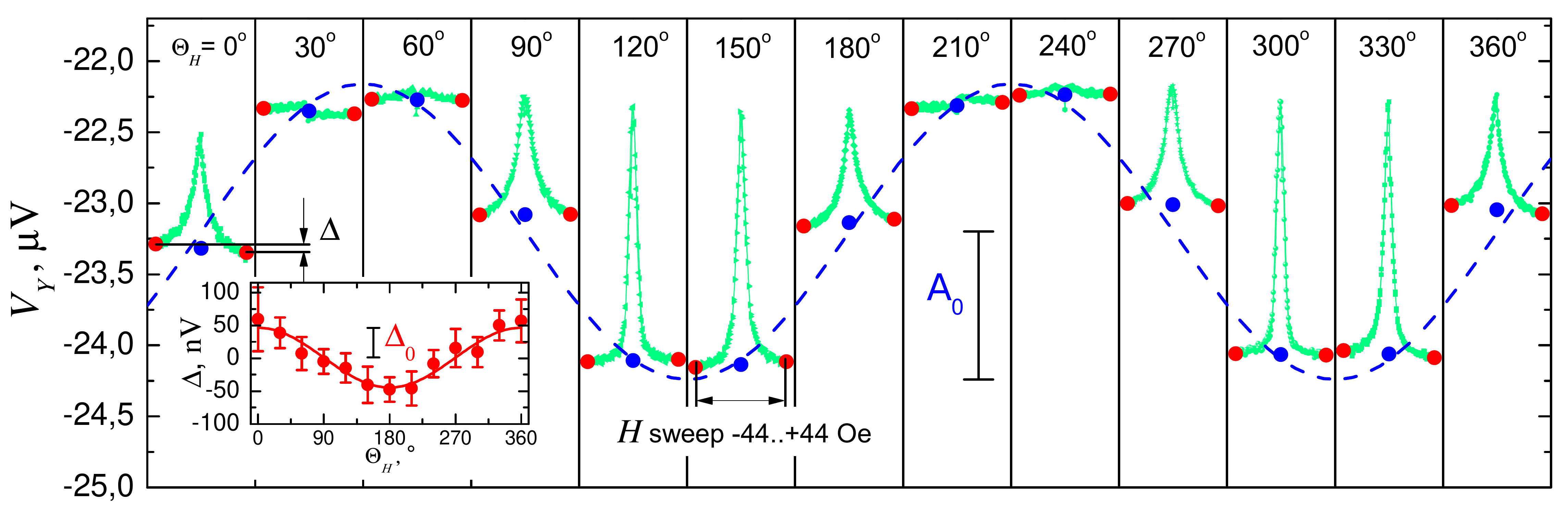}
		\caption{Set of transverse voltages $V_{y}(H)$ (green traces) for temperature difference $\Delta T_{x}=28$ K ($ \sim \nabla_{x} T=4.7$ K/mm), base temperature $T_{str}=360$ K and different $\Theta_{H}$. The average values $A$ (blue dots and blue dashed line) follow a $\sin2\Theta_{H}$ dependence with amplitude $A_{0}$ as expected from a contribution of the AMTEP. The difference values $\Delta$ (inset, red dots) shows a $\cos\Theta_{H}$ dependence with amplitude $\Delta_{0}$. This $\cos\Theta_{H}$ dependence could be associated with both TSSE and ANE.}
		\label{exmpl}
\end{figure*}

\section{Experimental data}

In fig.\,\ref{exmpl} we show a typical set of experimental data of the voltages $V_{y}(H)$ (green lines) measured across the Pt strip for the following parameters: base temperature $T_{str}=360$ K; temperature difference on the sample $\Delta T_{x}=28$ K ($ \sim \nabla_{x} T=4.7$ K/mm); sweeping field angles $\Theta_{H}=0^\circ,30^\circ,..,360^\circ$. The red dots shown on the left and right side of each green curve represent two averaged values of $V_{y}(H)$: for $H= -44..-40$ Oe and $ H=40..44$ Oe, respectively. From these two values we find the mean value $A(\Theta_{H})$ and the difference value $\Delta(\Theta_{H})$. $A(\Theta_{H})$ with magnitude $A_{0}$ is indicated by the blue dots and the blue dashed line, $\Delta(\Theta_{H})$ with magnitude $\Delta_{0}$ is shown in the inset.

The peaks visible in the $V_{y}(H)$ curves as well as the $\sin 2\Theta_{H}$ dependence of $A(\Theta_{H})$ are related to the AMTEP, for details see \cite{prlmax,AMTEP}. The transverse voltage, associated with the AMTEP traces, can be described as:
\begin{multline}
\textit {V}_{y}(H)-V_{offset}= A_{0}(\Delta T_x)\sin 2\Theta _{0}(H) \propto \\ \propto \left | \vec{M} \right |^2 \left | \vec{\nabla}_{x}T \right | \sin 2\Theta _{0}(H),
\label{AMTEP}
\end{multline}

where $\Theta_{0}$ is the angle between the temperature gradient $\vec{\nabla}_{x}T$ and the magnetization vector $\vec{M}$; $A_{0}$ is the AMTEP amplitude; $V_{offset}$ is the voltage signal produced in the Au wires due to the conventional charge Seebeck effect (for details see Appendix \ref{app:offsetv}). The peaks in $V_y(H)$ are caused by the alignment procedure of the magnetization $\vec{M}$ parallel to the magnetic easy axis of the Py film \cite{AMTEP}. The easy axis of the sample has an angle of $\varphi \approx 35^\circ$ with respect to the x-axis of the sample. For angles of the sweeping field $\Theta_{H}$ = $30^\circ,60^\circ,210^\circ,240^\circ$ we observe the smallest peaks in the AMTEP curves since the easy axis is close to the direction of the sweeping field $\vec{H}$. When the values of $H$ are large enough to saturate the sample, the magnetization direction angle $\Theta_{0}$ is very close to the angle $\Theta_{H}$ of the applied magnetic field. In this case the mean values $A$ should follow the $\sin 2\Theta_{H}$ dependence as can be seen for the blue dashed line.

The difference signal $\Delta(\Theta_{H})$ has a $\cos \Theta_{H}$ dependence (see inset in fig.\,\ref{exmpl}). Its amplitude $\Delta_{0}$ can be interpreted as a mixture of TSSE and ANE signals, where the ANE is produced by a spurious out-of-plane temperature gradient $\vec{\nabla}_{z} T$  \cite{prlmax}.

\begin{figure}
		\includegraphics[width=63mm]{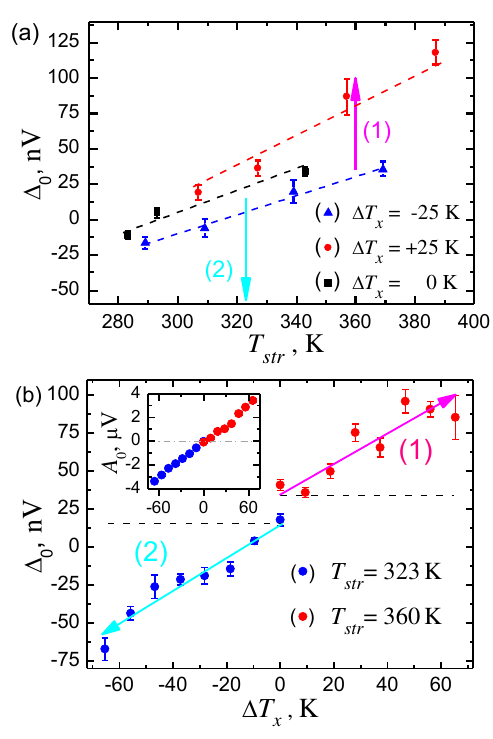} \\
		\caption{a) Dependence of the difference signal amplitude $\Delta_{0}$ on the base temperature $T_{str}$ for fixed temperature differences $\Delta T_{x}=-25, 0, 25$ K ($ \sim \vec{\nabla}_{x} T=-4.2, 0, 4.2$ K/mm respectively) taken from \cite{prlmax}. b) Dependence of the amplitude of the difference signal $\Delta_{0}$ on the temperature difference $\Delta T_{x}$ for fixed base temperatures $T_{str}=323$ K (blue dots and light blue arrow) and $360 $  K (red dots and magenta arrow) . The data shown in b) is represented in a) as the numbered arrows 1 (magenta) and 2 (light-blue). The inset represents the dependence of the AMTEP magnitude $A_0$ on $\Delta T_{x}$ and shows that at $\Delta T_x=0$ K $A_0$ is 0 V within the uncertainty range. This means that no in-plane temperature gradient is generated.}
		\label{exp}
\end{figure}

In fig.\,\ref{exp}a we replot the $\Delta_0$ experimental data that was published in \cite{prlmax}. These measurements have been performed for fixed temperature differences $\Delta T_{x}=-25, 0, +25$ K ($\nabla_{x} T=-4.2, 0, 4.2$ K/mm) and as a function of base temperature $T_{str}$. The slope of the $\Delta_{0}$ lines of fixed temperature differences $\Delta T_{x}=-25, 0, +25$ K (dashed lines in fig.\,\ref{exp}a) was explained in \cite{prlmax} as a contribution of ANE produced by a spurious $\vec{\nabla}_{z} T$ gradient due to surface-to-surface radiation heat transfer between the sample and the vacuum chamber ($\nabla_z T\propto (T_{room}-T_{str})^2$, where $T_{room}$ is the ambient temperature of the laboratory). At first sight, the shifts of the "undisturbed" line for $\Delta T_{x}=0$ K (up for a positive temperature gradient and down for a negative one) are expected to be related to TSSE. However, after having performed control measurements on samples where we have replaced the Pt detector strip with a Cu strip we found that the vertical shift is not related to the TSSE, but at this point its origin was an open question  \cite{prlmax}.

To clarify the nature of these shifts, we compare the data to additional measurements of $\Delta_0$ for a fixed $T_{str}$ and varying $\Delta T_{x}$ (see fig.\,\ref{exp}b). Here we use $T_{str}=360$ K for the strip being located at the hot side and $T_{str}=323$ K for it being located at the cold side. The inset in fig.\,\ref{exp}b proves that no in-plane temperature gradient is generated at the point $\Delta T_x=0$ K, since the AMTEP amplitude $A_0(\Delta T_x=0$ K) is equal to 0 V within the uncertainty range. Note that the black dashed lines in fig. \ref{exp}b indicate that $\Delta_{0}(\Delta T_{x}=0$ K) is not equal to 0 V for both $T_{str}=360$ K and $T_{str}=323$ K. This is caused by an ANE signal, since the base temperatures $T_{str}$ differs from $T_{room}$ (compare to the line for $\Delta T_{x}=0$ K in fig. \,\ref{exp}a). The experimental data shown in fig.\,\ref{exp}b is in addition represented in fig.\,\ref{exp}a as the numbered arrows. We note that the data matches the previously observed experimental findings: $\Delta_{0}$ increases with positive temperature gradients and decreases with negative temperature gradients.

\begin{figure}
		\includegraphics[width=46mm]{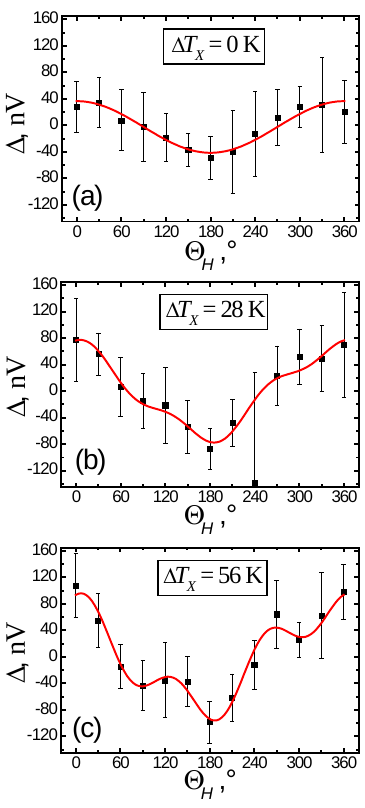}
		\caption{Difference signals $\Delta (\Theta_{H})$ for fixed base temperature  $T_{str}=360$ K and temperature differences: a) $\Delta T_{x}=0$ K ($\nabla_{x} T=0$ K/mm); b) $\Delta T_{x}=28$ K ($\nabla_{x} T=4.7$ K/mm); c) $\Delta T_{x}=56$ K ($\nabla_{x} T=9.5$ K/mm). With higher temperature differences $\Delta T_{x}$ a non cosine-like behavior appears. The red line is a fit function described in detail in the text.} .
		\label{Delta}
\end{figure}

Next we take a more precise look at the evolution of the $\Delta(\Theta_{H})$ traces as a function of $\Delta T_{x}$. Fig.\,\ref{Delta} reveals a new, non cosine-like contribution to the $\Delta(\Theta_{H})$ signal. This contribution becomes even more obvious for larger temperature gradients $\nabla_{x} T$. We will show in the following that it can be explained in terms of an antisymmetric AMTEP artifact due to extremely small static parasitic magnetic fields $\vec{H}_{p}$.

\section{Calculation of artifacts due to the AMTEP}
\begin{figure}
		\includegraphics[width=60mm]{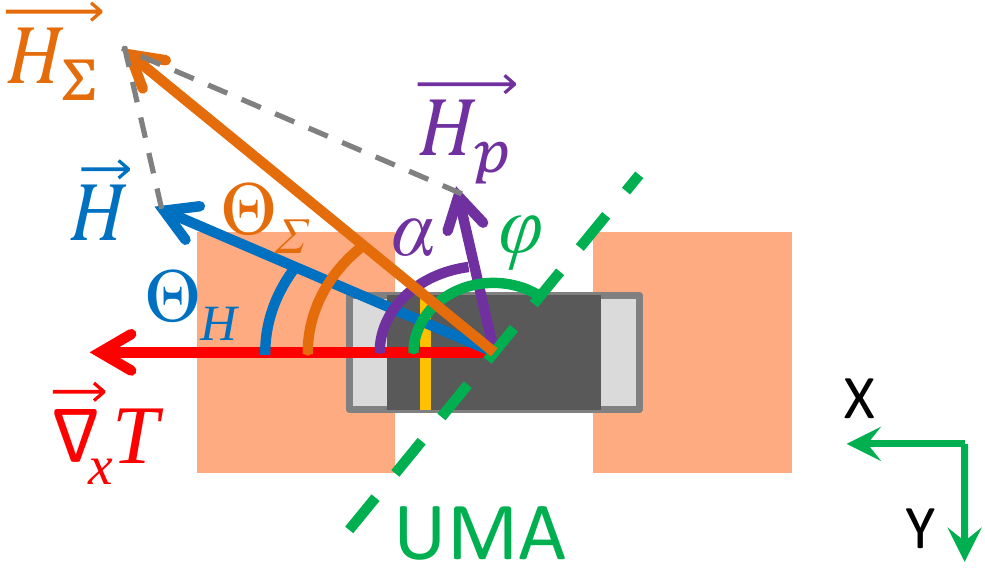}
		\caption{Definition of the coordinate system. The axis of the UMA is defined through the angle $\varphi$, the direction of the sweeping field $\vec{H}$ through the angle $\Theta_{H}$, the direction of the parasitic static magnetic field $\vec{H}_{p}$ through the angle $\alpha$ and the direction of the effective magnetic field $\vec{H}_{\Sigma}$ (vector sum of $\vec{H}$ and $\vec{H}_{p}$) through the angle $\Theta_{\Sigma}$. The temperature gradient $\vec{\nabla} T$ is parallel to the x-axis. }
		\label{Calcsetup}
\end{figure}

To simulate the AMTEP curves we need to know the equilibrium angle of the magnetization $\Theta_{0}$ (see eq.\,(\ref{AMTEP})) for every value of the sweeping field $H$. The angle $\Theta_{0}$ can be found by minimization of the magnetic free energy.
The model to calculate the free energy (see fig.\,\ref{Calcsetup}) includes an in-plane uniaxial magnetic anisotropy (UMA) with angle $\varphi$, sweeping field $\vec{H}$ with angle $\Theta_{H}$ and parasitic static magnetic field $\vec{H}_{p}$ with angle $\alpha$. The effective magnetic field $\vec{H}_{\Sigma}$ with angle $\Theta_{\Sigma}$ is found as vector sum of $\vec{H}$ and $\vec{H}_p$. All angles are counted with respect to the x-axis direction.

According to \cite{gurevich}, the in-plane density of magnetic free energy $U$ in the presence of an UMA reads:
\begin{multline}
U=-M_{0}H_{\Sigma}\cos(\Theta-\Theta_{\Sigma})+K\sin^{2}(\Theta-\varphi),
\label{energ}
\end{multline}
where the first term is a Zeeman energy, the second term represents the magnetocrystalline anisotropy energy, $M_{0}$ is the saturation magnetization,  $K$ is the constant describing the strength of the UMA and $\Theta$ is an arbitrary angle between the magnetization vector $\vec{M}$ and the temperature gradient $\vec{\nabla}_{x}T$.
In our model we exclude the demagnetization energy, since its in-plane contribution for our geometry is two orders of magnitude smaller than the energy of the magnetocrystalline anisotropy of our sample. We calculate the demagnetization factors according to \cite{demag} that leads to an effective in-plane demagnetization factor $(N_x-N_y)/4\pi$ of the order of $10^{-6}$ (effective in-plane demagnetizing field $\approx 0.01$ Oe).

The equilibrium angle $\Theta_{0}$ can now be found as a root of the derivative of the magnetic free energy $U$ (eq.\,(\ref{energ})) : 
			\begin{equation}
0=H_{\Sigma}\sin(\Theta_{0}-\Theta_{\Sigma})+\frac{H_{a}}{2}\sin 2(\Theta_{0}-\varphi),
\label{equil}
\end{equation}
where $H_{a}=\frac{2K}{M_{0}}$ is the UMA field. The subsequent calculation includes the following steps:

 %where effective field $H_\Sigma$ is amplitude of vector sum of $\vec{H}$ and $\vec{H_p}$, $\Theta_{\Sigma}$ is orientation angle of effective field $H_{\Sigma}$.

\begin{enumerate}
\item The parameters $\Theta_H, H_p, \alpha, H_a, \varphi$ are kept fixed. For each value of the sweep field ${H}$ (with a certain step in the range [$H_{max}$; $-H_{max}$], where $H_{max}$ is the maximum value of the sweep field) we compute the effective magnetic field $\vec{H}_{\Sigma}$ (with its effective angle $\Theta_{\Sigma}$) as a sum of the sweep field $\vec{H}$ and the parasitic field $\vec{H}_{p}$, see fig.\,\ref{Calcsetup}. After this step we know the dependencies $H_{\Sigma}(H)$ and $\Theta_{\Sigma}(H)$;
\item We solve eq.(\ref{equil}) numerically for the variable $\Theta_0$. This equation is solved for each value of $H$. After this step the dependence $\Theta_{0}(H)$ is known;
\item Inserting the computed dependence $\Theta_{0}(H)$, the experimental values $A_{0}(\Delta T_x)$ and $V_{offset}$ into eq.(\ref{AMTEP}) we build our calculated curve $V_{ycalc}(H)$, which we can compare with the experimentally determined $V_y(H)$.
%$V_{ycalc}(H)-V_{offset}$
\item We repeat steps 1.-3. for $\Theta_H=0^\circ..360^\circ$.
Finally, we calculate the AMTEP artifact difference signal $\Delta_{calc}(\Theta_H)=V_{ycalc}(\Theta_H, -H_{max})-V_{ycalc}(\Theta_H, H_{max})$.

\end{enumerate}

\begin{figure}
		\includegraphics[width=86mm]{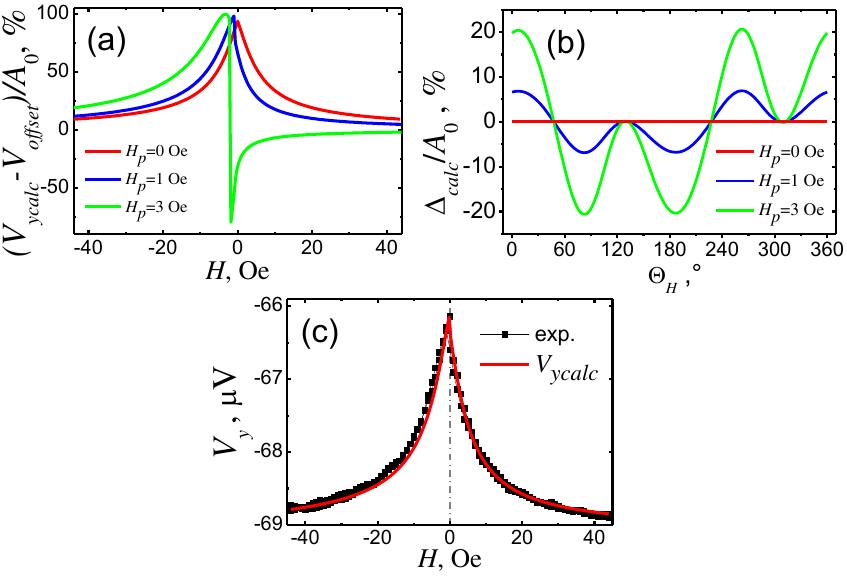}
		\caption{Appearance of an asymmetry in the AMTEP curves: a) Calculated AMTEP curves $(V_{ycalc}(H)-V_{offset})/A_0$ for $\Theta_{H}=0^\circ$ and $H_{p}=0, 1, 3$ Oe; b) Calculated difference signals $\Delta_{calc}(\Theta_H)$ related to $V_{ycalc}(H)$ from figure a; c) Experimental curve $V_y(H)$ for $T_{str}=360$ K, $\Delta T_{x}=65$ K ($ \sim \nabla_{x} T=11$ K/mm), $\Theta_{H}=0^\circ$ and calculated curve $V_{ycalc}(H)$ for $H_{p}=0.28$ Oe (the measured parasitic field in the current set-up).}
		\label{1curve}
\end{figure}

Fig.\,\ref {1curve}a shows that the calculated AMTEP curves $(V_{ycalc}(H)-V_{offset})$ shift and become asymmetric as the parasitic field $\vec{H}_{p}$ appears. The curves become actually antisymmetric for moderately high values of $H_{p}$ ($> 2$ Oe). We should note that a change of symmetry appears only if $\vec{H}_{p}\nparallel	 \vec{H}$, otherwise only a shift of the AMTEP curves can be observed. However, a difference signal $\Delta_{calc}(\Theta_{H})\neq 0$  caused by the parasitic field must be introduced, since $V_{ycalc}(H_{max})\neq V_{ycalc}(-H_{max})$ due to the induced shift. Fig.\,\ref {1curve}b presents the AMTEP artifact difference signals $\Delta_{calc}(\Theta_{H})$, related to the curves from fig.\,\ref {1curve}a. The artifact's magnitude obviously scales with the parasitic field value $H_p$ and in absence of a parasitic field the difference signal $\Delta_{calc}(\Theta_H)$ does not appear. 

%The calculated curves shown in fig.\,\ref {1curve}a are normalized by the AMTEP amplitude $A_{0}$.

\begin{figure} 
		\includegraphics[width=86mm]{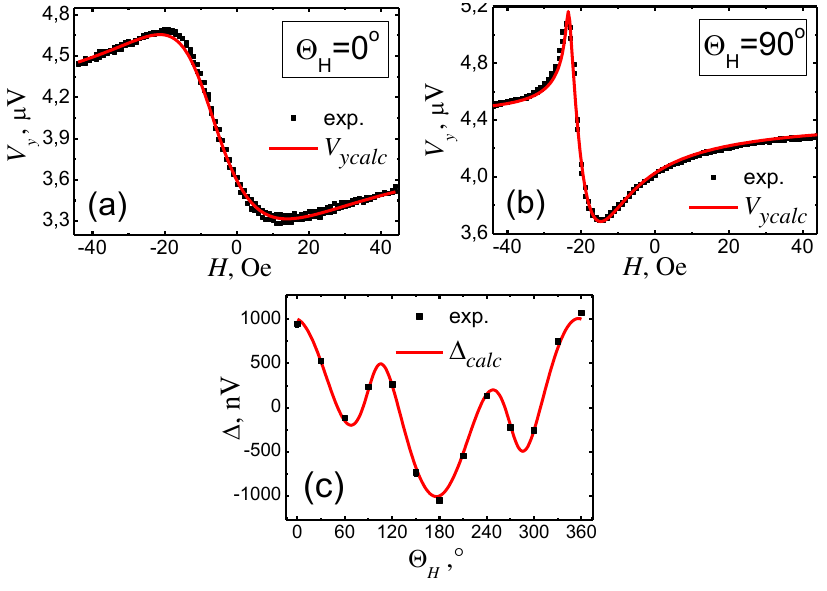}
		\caption{Experiments with large static parasitic magnetic field $H_{p}=19\pm 2$ Oe with angle $\alpha=-80\pm 3^\circ$ for $T_{str}=323$ K and $\Delta T_{x}=28 $ K ($ \sim \nabla_{x} T=4.7$ K/mm). Experimental $V_y(H)$ in comparison with calculated $V_{ycalc}(H)$:  a) Angle of the sweeping magnetic field $\Theta_{H}=0^\circ$; b) $\Theta_{H}=90^\circ$. c) Experimental $\Delta(\Theta_{H})$ in comparison with calculated artifact $\Delta_{calc}(\Theta_H)$. 
%		The calculated curves $V_{ycalc}(H)$ and $\Delta_{calc}(\Theta_H)$ perfectly reproduce the experimental curves $V_{y}(H)$ and $\Delta_(\Theta_H)$ respectively.
		}
		\label{Hpadd}
\end{figure}

The good match of the experimental data $V_y(H)$ with the calculation $V_{ycalc}(H)$ for $H_{p}=0.28$ Oe (the value of the parasitic field in our set-up) can be seen in fig.\,\ref{1curve}c. The parameters \cite{parms} of our set-up are used for all calculations in fig.\,\ref {1curve}.
To verify the accuracy of the computational model we perform measurements with large static parasitic magnetic field (several 10s of Oe, in this case AMTEP curves should be antisymmetric). A small permanent magnet was glued to the wall on the outside of the vacuum chamber and at the same height with the sample. The value and direction of the magnetic field at the position of the sample are determined with a Hall probe.
 One of the measurements for $H_{p}=19\pm 2$ Oe, $\alpha=-80\pm 3^\circ$, $T_{str}=323$ K and $\Delta T_{x}=28 $ K is shown in fig.\,\ref {Hpadd}. Fig.\,\ref {Hpadd}a and fig.\,\ref {Hpadd}b present a comparison of the calculated dependencies $V_{ycalc}(H)$ with the experimental $V_{y}(H)$ for angles of the sweeping field $\Theta_H=0^\circ$ and $90^\circ$. Fig.\,\ref {Hpadd}c, in turn, presents a comparison of the calculated dependence $\Delta_{calc}(\Theta_H)$ with the experimental $\Delta(\Theta_H)$. The calculated traces $V_{ycalc}(H)$ and $\Delta_{calc}(\Theta_H)$ nearly perfectly reproduce the experimental traces $V_{y}(H)$ and $\Delta(\Theta_H)$, respectively. A very small difference between $\Delta_{calc}(\Theta_H)$ and $\Delta(\Theta_H)$ due to contributions of ANE and TSSE signals to the experimental data are barely visible, since their magnitudes are by 2 orders smaller than artifact's magnitude.

% In fig.\,\ref {Hpadd}a ($\Theta_{H}=0^\circ$) we have higher mismatch of calculated and experimental data than in fig.\,\ref {Hpadd}a ($\Theta_{H}=0^\circ$). This is caused by the additional ANE contribution: for $\Theta_{H}=90^\circ$ it is equal to $0$ V (since the ANE has a $\cos\Theta_{H}$ dependence) and for $\Theta_{H}=0^\circ$ it has maximum value.

\begin{figure}
		\includegraphics[width=55mm]{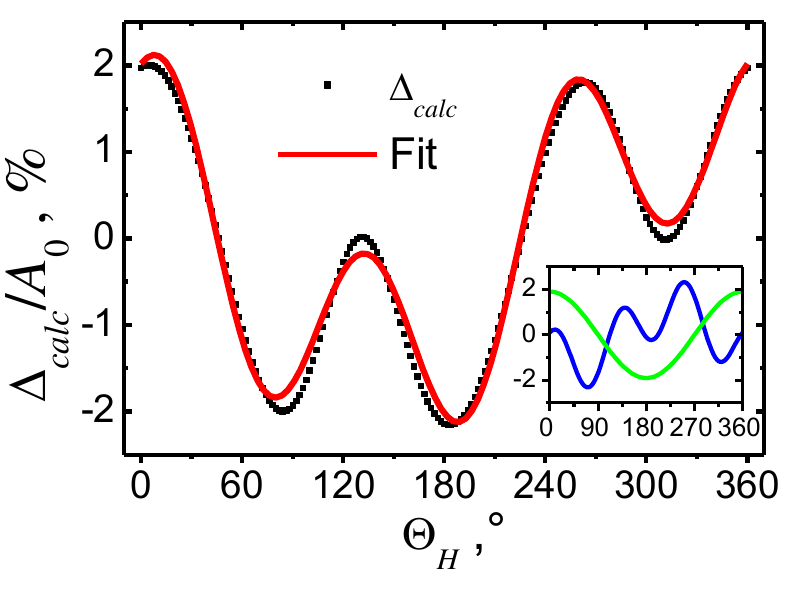} \\
		\caption{ $\Delta_{calc}(\Theta_{H})$ computed for the parameters of the experimental set-up \cite{parms}. The red line is a fit to equation (\ref{fit}). Artifacts of $\approx 2\%$ of the AMTEP amplitude $A_{0}$ are found. The inset presents the extracted fitted parts of the $\Delta_{calc}(\Theta_{H})$ signal: $\Delta_{AMTEP}\cos(\Theta_{H})$ (green line) and $E\sin(\Theta_{H}+\phi_1)\cos(2\Theta_{H}+\phi_2)$ (blue line). The fit curve deviates from the $\Delta_{calc}(\Theta_H)$ by less than $10\%$ of its amplitude, $R^{2}=0.99$.}
		\label{CalcD}
\end{figure}

In fig.\,\ref{CalcD} we present the calculated difference signal $\Delta_{calc}(\Theta_{H})$ (normalized by $A_{0}$) for the parameters \cite{parms} of the current experimental set-up.
$\Delta_{calc}(\Theta_{H})$ is in good agreement with the experimental curve $\Delta (\Theta_{H})$ shown in fig.\,\ref{Delta}c for a large temperature difference of $\Delta T_{x}=56$ K. The divergence of $\Delta(\Theta_{H})$ from $\Delta_{calc}(\Theta_{H})$ is caused by an additional cosine contribution of the ANE signal (with magnitude of $\approx35$ nV) to experimental data $\Delta(\Theta_{H})$.

Since we use numerical calculations, we cannot extract a functional dependence from $\Delta_{calc}(\Theta_{H})$. However, using a fit function:
\begin{multline}
\Delta(\Theta_{H})_{fit}=\Delta_{AMTEP}\cos \Theta_{H}+\\
+E\sin(\Theta_{H}+\phi_1)\cos(2\Theta_{H}+\phi_2)
\label{fit}
\end{multline}
we can match our calculations with less than $10\%$ error, $R^{2}=0.99$. Here $\Delta_{AMTEP} =4 A_{0}(\Delta T_x)\frac{H_{p}}{H_{max}}\sin(-\alpha)$ is the amplitude of the cos-like part of the artifact, where $H_{max}$ is the maximum value of the applied magnetic field; $E= 4 A_{0}(\Delta T_x)\frac{H_{p}}{H_{max}}$ is the amplitude of the non-cos-like part of the artifact; phase $\phi_1= 0^\circ$; phase $\phi_2= -\alpha$. The fit function (\ref{fit}) describes the behavior of the difference signal $\Delta_{calc}(\Theta_{H})$ very well until the conditions $\frac{H_p}{H_{max}}\ll 1$ and $\frac{H_a}{H_{max}}\ll 1$ are satisfied. The anisotropy field $H_a$ and its orientation angle $\varphi$ keep the shape of the calculated curve nearly undisturbed, until $\frac{H_a}{H_{max}}\ll 1$. The larger becomes $\frac{H_a}{H_{max}}$, the more the parameters $\Delta_{AMTEP}$, $E$, $\phi_1$ and $\phi_2$ deviate from the values defined above. After a certain point ($\frac{H_a}{H_{max}} \approx \frac{1}{2}$)  the difference signal $\Delta_{calc}(\Theta_{H})$ gets peaks, its shape changes strongly and it cannot be described with function (\ref{fit}) anymore. Angle $\alpha$ of a parasitic magnetic field strongly defines the phase $\phi_2$ and the amplitude $\Delta_{AMTEP}$: when $\vec{H}_{p}\parallel \vec{\nabla}_{x} T$ ($\alpha=0^\circ$) $\Delta_{AMTEP}= 0$ V; when $\vec{H}_{p}\perp \vec{\nabla}_{x} T$ ($\alpha=90^\circ$) $\Delta_{AMTEP}$ has a maximum value. When $\frac{H_p}{H_{max}}\gg 1$ the shape of the difference signal $\Delta_{calc}(\Theta_{H})$ transforms from function (\ref{fit}) into $-\sin(\Theta_{H}+\phi_2)$.  As shown in the inset of fig.\,\ref{CalcD}, artifact $\Delta_{calc}(\Theta_H)$ contains a significant cosine-like contribution that can be mistaken for a TSSE or ANE contribution.

The shape of the AMTEP artifact $\Delta_{calc}(\Theta_{H})$ does not depend on $\Delta T_{x}$, but its magnitude scales linearly with $\Delta T_{x}$. In this sense, the cosine-part of the artifact $\Delta_{AMTEP}$ has absolutely the same behavior as the TSSE. It shows the same angular dependence as TSSE and both effects are proportional to the temperature gradient $\nabla_{x} T$. The ANE, in turn, shares only the angular dependence with the TSSE.
	
Since we found a new fit function (\ref{fit}), we can use it to fit the experimental difference signals $\Delta(\Theta_{H})$. We only replace $\Delta_{AMTEP}$ in (\ref{fit}) by $\Delta_{0}$, which in case of experimental data contains three contributions:
\begin{multline}
\Delta_{0}(T_{str}, \Delta T_x)=\overbrace{\Delta_{ANE}(T_{str})}^{\propto T_{str}^2}+ \\+\overbrace{\Delta_{AMTEP}(\Delta T_{x})+\Delta_{TSSE}(\Delta T_{x})}^{\propto \Delta T_x},
\label{expD}
\end{multline}
where $\Delta_{AMTEP}$ is the cosine-part of the AMTEP artifact, $\Delta_{ANE}$ the contribution of the ANE due to a spurious out-of-plane temperature gradient $\vec{\nabla}_{z}T$ and $\Delta_{TSSE}$ the contribution of the TSSE. The phases $\phi_1$ and $\phi_2$ in (\ref{fit}) are extracted from the $\Delta_{calc}(\Theta_{H})$ in fig.\,\ref{CalcD} (which computed for actual set-up parameters \cite{parms}) and are then kept constant in (\ref{fit}) for fitting the experimental data.
	
\section{Interpretation of the experimental results }

\begin{figure}
		\includegraphics[width=60mm]{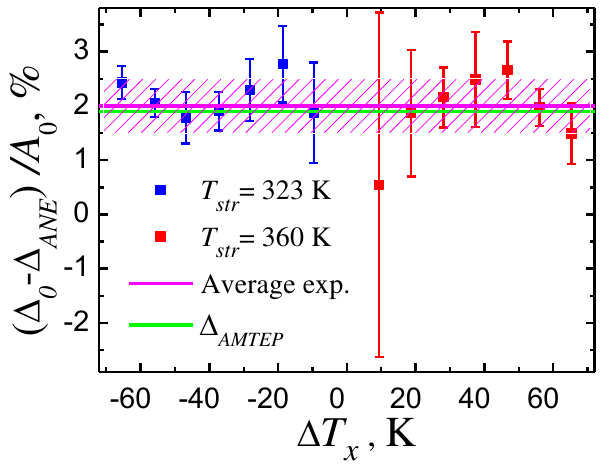}
		\caption{Comparison of the experimental level $(\Delta_0-\Delta_{ANE})/A_0$  (blue and red points, magenta line with shaded area presents averaged experimental data) with the normalized calculated AMTEP artifact $\Delta_{AMTEP}/A_{0}$ (green line). Large deviations of the experimental points from the average value in the region of small differences $\Delta T_{x}$ are connected to small values of the $\Delta_{0}-\Delta_{ANE}$ and are proportionally impacted by noise.}
		\label{DE}
\end{figure}

The vertical shift of the line for $\Delta T_{x}=0$ K seen in fig.\,\ref{exp}a , when changing the temperature difference $\Delta T_{x}$, is mostly caused by the AMTEP artifact. It can finally be proven by showing the data in fig.\,\ref{DE}. The experimental data points are calculated by subtracting the ANE contribution $\Delta _{ANE}=\Delta _{0}(\Delta T_{x}=0$ K) from the experimental amplitude $\Delta _{0}(\Delta T_{x})$, see eq.\,(\ref{expD}) (i.e. this points contain mixture of TSSE signal magnitude $\Delta_{TSSE}$ and cosine-part of AMTEP artifact $\Delta_{AMTEP}$). The normalized amplitude of the calculated cosine-part of the AMTEP artifact amounts to $\Delta _{AMTEP}/A_{0}$ = 1.9 $\%$ (green line). This value is in good agreement with the experimental value ($\Delta_0-\Delta_{ANE}$)/$A_0$ of $2 \pm 0.5\%$ (magenta line and shaded area). This means, that the maximum possible contribution of the TSSE inside the error bars is  $<\pm$ 0.3 nV/K ($\sim$ $\pm2\cdot10^{-12} $\;V$\cdot$m/K in terms of temperature gradient), which leads to $\pm20$ nV for the maximum achieved temperature differences $\Delta T_{x}=\pm 65$ K.

We have examined possible sources of parasitic magnetic fields in our set-up and found that a parasitic field $H_{p}=0.28$ Oe is caused by external magnetic fields existent in the laboratory. We have also examined the magnetic field contribution from the Peltier elements and verified that it produces only an insignificant field of 0.03 Oe maximum.

There are few possibilities to suppress the cos-part of the AMTEP artifact:
\begin{enumerate}
\item Increasing of the maximum value of the sweep field $H_{max}$, as the equilibrium angle $\Theta_{0}$ will then be closer to the angle of the sweep field $\Theta_{H}$. According to our calculations, the level of the artifact could be reduced from 2\% of the AMTEP amplitude $A_{0}$ (equal to 70 nV for the highest temperature gradients $\nabla_{x} T=\pm 11$ K/mm) to $0.06\%$ (equivalent to 2 nV $\ll$ than noise level of 20 nV) by increasing the $H_{max}$ from $44$ Oe to $2000$ Oe. The level of the artifacts is inversely proportional to the $H_{max}$. Unfortunately, in our setup we have no possibility to produce high magnetic fields larger than $50$ Oe.
\item Keeping the sweep field $\vec{H}$ parallel to the parasitic field $\vec{H}_{p}$ and rotating the sample together with the applied in-plane temperature gradient $\vec{\nabla} T$. In this case the symmetry of the AMTEP traces is not disturbed, but a horizontal shift of traces equal to the value $H_p$ appear. After this shift is subtracted, no AMTEP artifact in the difference signal $\Delta(\Theta_H)$ will be observed. However, rotation of the vacuum chamber is in our case not feasible.
\item Keeping the parasitic field $\vec{H}_{p}$ parallel to the applied in-plane temperature gradient $\vec{\nabla} T$. With this alignment the cos-part of AMTEP artifact would be strongly suppressed. The lower is the ratio $\frac{H_a}{H_{max}}$ and the closer is $\varphi$ to $n\cdot90^\circ$ (where n is integer), the stronger is the suppression. However, the non-cos-part of AMTEP artifact will be not suppressed.
\item The best solution is to suppress $H_p$ using additional compensating magnetic field.
\end{enumerate}

\section{Conclusion}
In this study we have shown that a small parasitic static magnetic field below 1 Oe can produce an AMTEP artifact with amplitude of 10-100 nV that contains a TSSE-like contribution. This artifact has the same order of magnitude as the previously found source of misinterpretations in the TSSE experiments - the ANE due to out-of-plane temperature gradient (produced by heat flux through electrical contacts \cite{AMTEP} and by heat radiation to the surrounding area \cite{prlmax}). But in contrast to the ANE-based parasitic effects, the new artifact does depend on the applied in-plane temperature gradient. This fact makes it more difficult to separate from possible TSSE signal, and, consequently, it has to be seriously taken into account. We found that the investigated parasitic effect also gives us the missing explanation of the vertical shift observed in \cite{prlmax}. Finally, no TSSE signals larger than the noise level of 20 nV have been detected in permalloy thin films.

\section*{Acknowledgments}
We acknowledge financial support by the Deutsche Forschungsgemeinschaft (DFG) through SPP 1538. The research leading to these results has received funding from the European Union Seventh Framework Programme (FP7-People-2012-ITN) under grant agreement 316657 (SpinIcur).

\appendix
%\begin{appendicies}
%\addappheadtotoc
\numberwithin{equation}{section} 
\numberwithin{figure}{section} 

\section{\MakeUppercase{Temperature measurements}} \label{app:temp}

\begin{figure} [h]
	\centering
		\includegraphics[width=8cm]{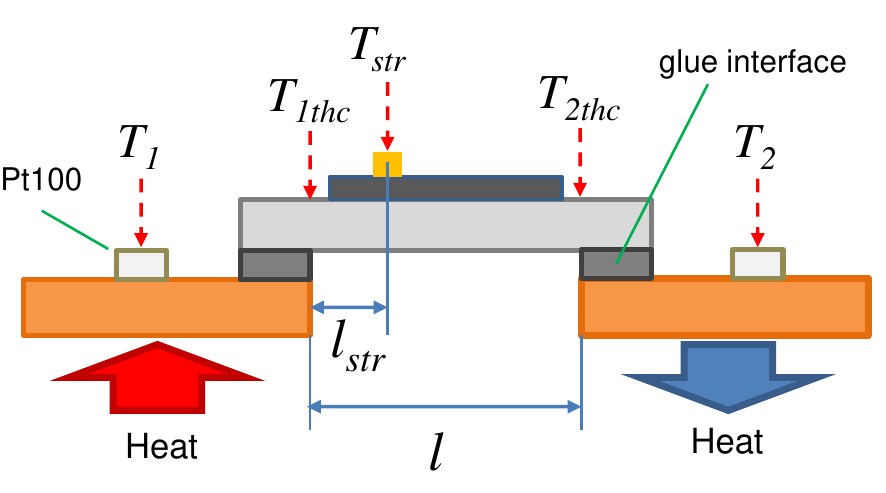}
		\caption{Definition of certain temperatures: $T_1$ and $T_2$ are temperatures of Peltier elements measured by temperature sensors Pt100; $T_{str}$ is temperature at position of Pt strip; $T_{1thc}$ and $T_{2thc}$ are temperatures at the edges of suspended part of sample.}
		\label{tdef}
\end{figure}

The effective temperature difference on the sample $\Delta T_{x}$ can be evaluated:
\begin{equation}
\Delta T_{x}=(1-2\alpha_t)(T_{1}-T_{2}),
\label{dtx}
\end{equation}

where $T_{1}$ is the temperature measured with the Pt100 sensor on the Cu interface of the Peltier element that is closer to the Pt strip of the sample (see fig.\,\ref{tdef} - left heat sink with Pt 100 sensor on it), $T_{2}$ is the temperature measured on the second Peltier element, $\alpha_t$ is a dimensionless coefficient that describes temperature "losses" $\alpha_t (T_1-T_2)$ on each of the glue interfaces between Peltier elements and sample. To find $\alpha_t$ we conduct measurements with 2 additional thermocouples of type K. Thermocouples are placed on top of the sample in the points $T_{1thc}$ and $T_{2thc}$ shown of fig.\,\ref{tdef}. Difference between temperatures measured by these thermocouples gives us $\Delta T_{x}=T_{1thc}-T_{2thc}$. We assume that temperature changes linearly with distance from point with temperature $T_{1thc}$ to point with temperature $T_{2thc}$. Thus:
\begin{equation}
T_{str}=T_{1thc}-\Delta T_{x}\frac{l_{str}}{l},
\label{dtx}
\end{equation}

where $l_{str}$ is the distance from point with $T_{1thc}$ to the position of the Pt strip, $l$ - distance between point with $T_{1thc}$ and point with $T_{2thc}$. Since we fix certain $T_{str}$ and $\Delta T_{x}$ in our measurements, we need to know at which temperatures both Peltier elements should be kept. These temperatures can be calculated as follows:

\begin{equation}
\left.\begin{aligned}
T_{1}=T_{str}+\Delta T_{x}(\frac{l_{str}}{l}+\frac{\alpha_t}{1-2\alpha_t})\\
T_{2}=T_{str}-\Delta T_{x}(\frac{l-l_{str}}{l}+\frac{\alpha_t}{1-2\alpha_t})
\end{aligned} \right\}
\label{T12}
\end{equation}

\section{\MakeUppercase{Offset voltage in measurements}} \label{app:offsetv}

\begin{figure} [h]
	\centering
		\includegraphics[width=8cm]{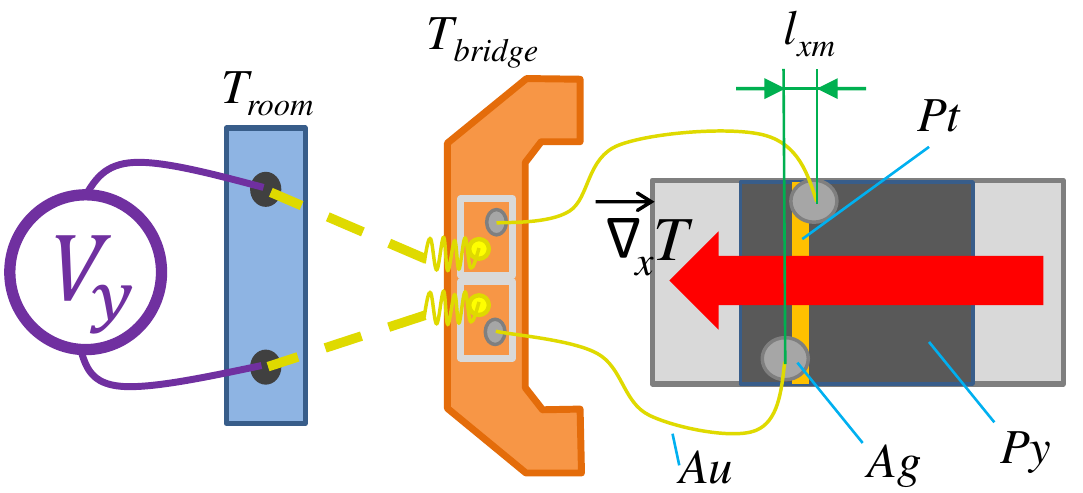}
		\caption{Appearance of thermoelectric signal in Au wires. Ends of Au wires those connected to voltmeter are kept at temperature $T_{bridge}$ which is close to room temperature $T_{room}$. Opposite ends are glued to Pt strip. Small x-coordinate mismatch $l_{xm}$ of centers of glue drops in presence of temperature gradient $\vec{\nabla}_x T$ lead to temperature mismatch of these ends of Au wires.}
		\label{voff}
\end{figure}

The functional dependence of $V_{offset}$ is given by: 

\begin{equation}
V_{offset} = \alpha_{1} \Delta T_{x} + \alpha_{2} \Delta T_{x}(T_{str}-T_{room}),
\label{Vof}
\end{equation}

where $T_{room}$ is room temperature, $\alpha_{1}$ and $\alpha_{2}$ are the effective first and second order coefficients of the conventional charge Seebeck effect. We found in our measurements $T_{room}\approx 296$ K, $\alpha_{1}=-0.53$ $\mathrm{\mu V/K}$ and  $\alpha_{2}=-4.6$ nV/K$^{2}$. In this case the coefficients $\alpha_{1}$ and $\alpha_{2}$ are not only usual thermoelectric coefficients, but thermoelectric coefficients of Py/Ag interface multiplied by the geometrical mismatch coefficient $\frac{l_{xm}}{l}$, where $l_{xm}$ is the x-projection of the distance between the centers of the glue drops that connect the Au wires with the Pt strip (see fig.\,\ref{voff}), $l$ is the length of the suspended part of the sample (see fig.\,\ref{tdef}). We consider a Py/Ag interface since Ag is the main component of the glue that connects the Au wires with the Pt strip of the sample and since the glue spots are distributed not only over the Pt strip, but also over the Py film. We take into account only the Py/Ag interfaces for the estimation of the thermoelectric coefficients since the Pt/Ag interfaces stay at the same temperature and produces no thermoelectric voltage. The thermoelectric voltage produced by the Au/Ag interface is 2 orders of magnitude smaller than the voltage produced by the Py/Ag interface. All other pairs of electrical interfaces, starting from the Cu bridge and finishing by the direct contacts of the nanovoltmeter, do not produce the thermoelectric voltage, since all pairs stay at the same temperatures.  

%\end{appendicies}


\begin{thebibliography}{000}

%\bibitem[]{spcal1} G. E.W. Bauer, A.H. MacDonald, and S. Maekawa \textit{Solid State Commun.} \textbf{150}, 459 (2010).

%\bibitem[]{spcal2} M. Johnson \textit{Solid State Commun.} \textbf{150}, 543-547 (2010).

\bibitem[]{timo1} G. E.W. Bauer, E. Saitoh, and B. J. van Wees \textit{Nat. Mater.} \textbf{11}, 391 (2012).

\bibitem[]{timo2} S. R. Boona, R. C. Myers, and J. P. Heremans \textit{Energy Environ. Sci.} \textbf{7}, 885 (2014).

%\bibitem[]{lsse1} K. Uchida, H. Adachi, T. Ota, H. Nakayama, S. Maekawa, and E. Saitoh \textit{Appl. Phys. Lett.} \textbf{97}, 172505 (2010).

%\bibitem[]{lsse2} D. Meier, T. Kuschel, L. Shen, A. Gupta, T. Kikkawa, K.Uchida, E. Saitoh, J.-M. Schmalhorst, and G. Reiss \textit{Phys.Rev. B} \textbf{87}, 054421 (2013).

\bibitem[]{tsse1} K. Uchida, S. Takahashi, K. Harii, J. Ieda,W. Koshibae, A.Ando, S. Maekawa, and E. Saitoh \textit{Nature (London)} \textbf{455}, 778 (2008).

\bibitem[]{tsse2} A. D. Avery, M.R. Pufall, and B. L. Zink \textit{Phys. Rev. Lett.} \textbf{109}, 196602 (2012).

\bibitem[]{tsse3} K. Uchida, T. Ota, K. Harii, K. Ando, H. Nakayama, and E. Saitoh \textit{J. Appl. Phys.} \textbf{107}, 09A951 (2010).

\bibitem[]{tsse4} S.Y. Huang, W. G. Wang, S. F. Lee, J. Kwo, and C. L. Chien \textit{Phys. Rev. Lett.} \textbf{107}, 216604 (2011).

\bibitem[]{tsse5} S. Bosu, Y. Sakuraba, K. Uchida, K. Saito, T. Ota, E. Saitoh, and K. Takanashi \textit{Phys. Rev. B} \textbf{83}, 224401 (2011).

\bibitem[]{tsse6} C. M. Jaworski, J. Yang, S. Mack, D. D. Awschalom, J. P. Heremans, and R. C. Myers \textit{Nat. Mater.} \textbf{9}, 898 (2010).

\bibitem[]{tsse7} C. M. Jaworski, J. Yang, S. Mack, D. D. Awschalom, R. C. Myers, and J. P. Heremans \textit{Phys. Rev. Lett.} \textbf{106}, 186601 (2011).

\bibitem[]{tsse8} K. Uchida, J. Xiao, H. Adachi, J. Ohe, S. Takahashi, J. Ieda, T. Ota, Y. Kajiwara, H. Umezawa, H. Kawai, GE.W. Bauer, S. Maekawa, and E. Saitoh \textit{Nat. Mater.} \textbf{9}, 894 (2010).

\bibitem[]{tsse9} H. Adachi, K. Uchida, E. Saitoh, J. Ohe, S. Takahashi, and S. Maekawa \textit{Appl. Phys. Lett.} \textbf{97}, 252506 (2010).

\bibitem[]{timo3} I. V. Soldatov, N. Panarina, C. Hess, L. Schultz, and R. Sch{\"a}fer \textit{Phys. Rev. B} \textbf{90}, 104423 (2014).

\bibitem[]{timo4} C. T. Bui and F. Rivadulla \textit{Phys. Rev. B} \textbf{90}, 100403 (2014).

\bibitem[]{timo5} D. Meier, D. Reinhardt, M. van Straaten, C. Klewe, M. Althammer, M. Schreier, S. T. B. Goennenwein, A. Gupta, M. Schmid, C. H. Back, J.-M. Schmalhorst, T. Kuschel, and G. Reiss \textit{Nat. Commun.} \textbf{6}, 8211 (2015).

\bibitem[]{prlmax} M. Schmid, S. Srichandan, D. Meier, T. Kuschel, J.-M. Schmalhorst, M. Vogel, G. Reiss, C. Strunk and C.H. Back \textit{Phys. Rev. Lett.} \textbf{111}, 187201 (2013).

\bibitem[]{AMTEP} D. Meier, D. Reinhardt, M. Schmid, C.H. Back, J.-M. Schmalhorst, T. Kuschel and G. Reiss \textit{Phys. Rev. B} \textbf{88}, 184425 (2013).

\bibitem[]{timoANE} A. von Ettingshausen and W. Nernst \textit{Annu. Rev. Phys. Chem.} \textbf{265}, 343 (1886).

\bibitem[]{AMTEP2} Yong Pu, E. Johnston-Halperin, D.D. Awschalom and Jing Shi \textit{Phys. Rev. Lett.} \textbf{97}, 036601 (2006).

\bibitem[]{AMTEP3} V.D. Ky \textit{Phys. Status Solidi} \textbf{22}, 729 (1967).

\bibitem[]{ishe1} M.I. D'yakonov and V.I. Perel \textit{Phys. Lett.} \textbf{35}, 459 (1971).

\bibitem[]{timo6} E. Saitoh, M. Ueda, H. Miyajima, and G. Tatara \textit{Appl. Phys. Lett.} \textbf{88}, 182509 (2006).

\bibitem[]{ishe2} T. Kimura, Y. Otani, T. Sato, S. Takahashi, and S. Maekawa \textit{Phys. Rev. Lett.} \textbf{98}, 156601 (2007).

%\bibitem[]{ishe3} O. d'Allivy Kelly, A. Anane, R. Bernard, J. Ben Youssef, C. Hahn, A H. Molpeceres, C. Carrétéro, E. Jacquet, C. Deranlot, P. Bortolotti, R. Lebourgeois, J.-C. Mage, G. de Loubens, O. Klein, V. Cros, and A. Fert \textit{Appl. Phys. Lett.} \textbf{103}, 082408 (2013).

\bibitem[]{gurevich} A.G. Gurevich, G.A. Melkov, \textit{Magnetization oscillations and waves}, CRC Press, Inc. (1996).

\bibitem[]{demag} Amikam Aharoni \textit{J. Appl. Phys.} \textbf{83}, 3432 (1998).

\bibitem[]{parms} $H_{p}=0.28$ Oe, $\alpha=-51^\circ$, $H_{a}=4.5$ Oe, $\varphi=35^\circ$, $H=-44..+44$ Oe. Since we perform averaging of the experimental $V_{y}$ data points from $\pm 40$ to $\pm 44$ Oe, a range of $H=-42..+42$ is used for the simulation of $\Delta_{calc} (\Theta_{H})$.


\end{thebibliography}
\end{document}